\documentclass[notitlepage,superscriptaddress,showpacs,a4paper,nobalancelastpage,prl,twocolumn]{revtex4-1}%
\usepackage{amsfonts}
\usepackage{subfigure}
\usepackage{amsmath}
\usepackage{amssymb}
\usepackage{graphicx,epstopdf}
\usepackage{appendix}
\usepackage{array}
\usepackage{color}
\usepackage{my_symbols}
\usepackage{tikz}
\usetikzlibrary{calc,3d}
\usetikzlibrary{arrows}
\usepackage[normalem]{ulem}
\usepackage{float}
\providecommand{\U}[1]{\protect\rule{.1in}{.1in}}

\newcommand\rmv{\bgroup\markoverwith {\textcolor{red}{\rule[0.5ex]{2pt}{0.4pt}}}\ULon}

\begin{document}
\title{Spin Hall effect by surface roughness}
\author{Lingjun Zhou}
\fudan
\author{Vahram L. Grigoryan}
\fudan
\author{Sadamichi Maekawa}
\affiliation{Advanced Science Research Center, Japan Atomic Energy Agency, Tokai 319-1195, Japan}
\affiliation{CREST, Japan Science and Technology Agency, Tokyo 102-0075, Japan}
\author{Xuhui Wang}
\email[Corresponding author:~]{xuhui.wang@kaust.edu.sa}
\affiliation{King Abdullah University of Science and Technology (KAUST), Physical Science and Engineering Division, Thuwal 23955-6900, Saudi Arabia}
\author{Jiang Xiao}
\email[Corresponding author:~]{xiaojiang@fudan.edu.cn}
\fudan
\begin{abstract}

The spin Hall effect and its inverse effect, caused by the spin orbit interaction, provide the interconversion between spin current and charge current. Since the effects make it possible to generate and manipulate spin current electrically, how to realize the large effects is an important issue in both physics and applications.  To do so, materials with heavy elements, which have strong spin orbit interaction, have been examined so far.  Here, we propose a new mechanism to enhance the spin Hall effect without heavy elements, \ie surface roughness in metallic thin films. We examine Cu and Al thin films with surface roughness and find that they give the spin Hall effect comparable to that in bulk Au. We demonstrate that the spin Hall effect induced by surface roughness has the side jump contribution but not skew scattering.

\end{abstract}

\maketitle


The spin Hall effect, caused by the spin orbit interaction, converts a charge current into a pure spin current  \cite{karplus_hall_1954,dyakonov_current-induced_1971,dyakonov_possibility_1971,hirsch_spin_1999,sinova_universal_2004,kato_observation_2004,wunderlich_experimental_2005,valenzuela_direct_2006,kimura_room-temperature_2007,seki_giant_2008,takahashi_spin_2008}. Its inverse effect \cite{saitoh_conversion_2006}, the inverse spin Hall effect converts a spin current into a charge current. These two effects make it possible to generate and detect spin current electrically. There are many applications of spin Hall effect and inverse spin Hall effect in magnetization switching, domain wall motion, spin current detection \etc. All such applications requires high efficiency of spin and charge conversion, therefore how to realize large spin Hall effect is an important research topic. 

There are different mechanisms for the spin Hall effect: intrinsic mechanism \cite{karplus_hall_1954,murakami_dissipationless_2003,sinova_universal_2004,inoue_suppression_2004,wunderlich_experimental_2005,zhang_intrinsic_2005,tse_spin_2006,krotkov_intrinsic_2006} from band structure properties and extrinsic mechanism from impurity scattering. Two processes contribute to the extrinsic mechanism: side jump and skew scattering \cite{mott_theory_1949,smit_spontaneous_1958,berger_side-jump_1970,dyakonov_current-induced_1971,nozieres_simple_1973,zhang_spin_2000,tse_spin_2006,takahashi_spin_2008}, where the former comes from lateral displacement of the wave function during the scattering event and the latter comes from the spin dependence of scattering cross section. Both extrinsic and intrinsic spin Hall effect requires large spin orbit interaction, which usually exists only in heavy atoms. So spin Hall effect is typically linked with materials consist of heavy elements, for example noble metals: Au, Pt, Ta. There has been a lot of attempts to use different elements or different compositions to realizing large spin Hall effect. In order to achieve large spin Hall effect, it is helpful to explore other possibilities of creating spin Hall effect beyond simply trying heavy elements. A recent attempt in this front was to use normal metal thin films (such as Cu, Al), and the inversion symmetry breaking at the film surfaces induces interfacial Rasbha spin orbit interaction and gives rise to intrinsic spin Hall effect \cite{wang_spin-hall_2013}.

In this Letter, we propose a new strategy in realizing extrinsic spin Hall effect by using the scattering from rough surface in metallic thin films. The rough surface can be regarded as impurities. The finite thickness of the thin film causes the discrete energy in the thickness direction due to the confinement. The discrete energy levels depend on the film thickness, and at locations where the thickness is thinner (thicker) than the average thickness the confinement can be regarded as repulsive (attractive) potential hill (valley). Thus electrons transporting within the film are being scattered constantly in this random potential landscape. The associated spin orbit interaction of the random potential gives rise to the spin Hall effect with side jump mechanism, but no skew scattering.


\begin{figure}[b]
\begin{tikzpicture}
\draw[->] (-4.2,-0.3) -- (-3.7,-0.3) node[right] {$x$};
\draw[->] (-4.3,-0.2) -- (-4.3,0.3) node[above] {$z$};
\node[left] at (-4.35,-0.3) {$y$};
\node at (-4.3,-0.3) {${\otimes}$};

\pgfmathsetseed{2}
\coordinate (current point) at (-4,0.2*rand);
\foreach \i in {1,...,35} { 
\filldraw[orange!30,thick] (current point) --  (0.2*\i-4,0.2*rand) coordinate (current point) -- (0.2*\i-4,1) -- (0.2*\i-4.2,1); }

\pgfmathsetseed{2}
\coordinate (current point) at (-4,0.2*rand);
\foreach \i in {1,...,35} { 
\draw[thick] (current point) --  (0.2*\i-4,0.2*rand) coordinate (current point); }

\pgfmathsetseed{3}
\coordinate (current point) at (-4,2+0.2*rand);
\foreach \i in {1,...,35} { 
\filldraw[orange!30,thick] (current point) --  (0.2*\i-4,2+0.2*rand) coordinate (current point) -- (0.2*\i-4,1) -- (0.2*\i-4.2,1); }

\pgfmathsetseed{3}
\coordinate (current point) at (-4,2+0.2*rand);
\foreach \i in {1,...,35} { 
\draw[thick] (current point) --  (0.2*\i-4,2+0.2*rand) coordinate (current point); }
\draw[|<->|] (3.2,0) -- node[right] {$d$} (3.2,2);
\draw[thick,dashed,red] (-4,0) -- (3,0);
\draw[thick,dashed,red] (-4,2) -- (3,2);

\draw[thick,->,blue] (-1.5,1) node[below] {$\ket{n\bq s}$} -- (-0.24,1.85) -- (0.5,1) node[below] {$\ket{n'\bq's'}$};
\end{tikzpicture}
\caption{(Color online) A metallic thin film with rough surfaces, the film thickness at $\brho = (x,y)$ is $d(\brho)$ with $\avg{d(\brho)} = d$. }
\label{film}
\end{figure}
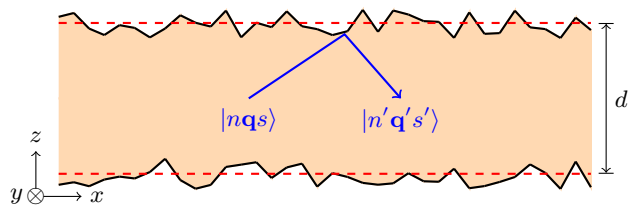

We consider a normal metallic thin film with rough surfaces as shown in \Figure{film}, where the film is confined in $z$ direction and extends in $\brho = (x,y)$ direction. The film thickness is position dependent $d(\brho)$ with $\avg{d(\brho)} = d$ being the average thickness, $\avg{\cdots}$ denotes the ensemble average. Let $\bp_\|$ and $p_z$ be the momentum operator in the $\brho$ and $z$ directions, then the Hamiltonian is 
\begin{equation}
H = {\bp_\|^2\ov 2m^*} + \midb{{p_z^2\ov 2m^*} + V_{d(\brho)}(z)} = H_\| + H_\perp^{d(\brho)},
\end{equation}
where $H_\perp^{d(\brho)}$ describes the confined quantum well states in $z$ direction, $m^*$ is the electron effective mass, and $V_{d(\brho)}(z)$ is the confining potential with the variable length scale $d(\brho)$. The most convenient way of handling surface roughness is to use the dilation operator. When $d(\brho)$ only slightly deviates from its average value $d$, the dilation operator \cite{tesanovic_quantum_1986,trivedi_quantum_1988}
$U = \exp\smlb{\lambda_{\brho}}\exp\midb{\lambda_{\brho}(z\partial_z + \partial_z z)/2}$
with $\lambda_{\brho} \equiv \ln[d/d(\brho)]$ dilates $H_\perp^d$ for a quantum well with constant thickness $d$ into $H_\perp^{d(\brho)}$ for a well with variable thickness: 
$H_\perp^{d(\brho)} = U H_\perp^d U^\dagger = H_\perp^d + V_R$
with $V_R = \lambda_{\brho}\smlb{2V_d + z\partial_z V_d}$
is an effective surface scattering potential that takes the full responsibility of the surface roughness. For simplicity, we consider the 'white noise' surface profile, \ie the surface roughness is uncorrelated and characterized by the dimensionless variance parameter $\Lambda \sim (\delta/d)^2$ with the thickness deviation variance $\delta^2$, so the correlation $\Avg{\lambda_{\brho}\lambda_{\brho'}} = \Lambda a^2\delta(\brho-\brho')$ with the lattice constant $a \sim k_F^{-1}$. 

Now, we introduce two new terms: i) the potential due to the bulk impurities: $V_I = V_{\rm imp}k_F^{-3} \sum_i\delta(\brho-\brho_i) \delta(z-z_i)$, where $(\brho_i,z_i)$ is the position of impurity-$i$ and $V_{\rm imp}$ is the magnitude of the $\delta$-like impurity potential, ii) the spin orbit interaction due to the surface scattering potential $V_R$: $V_R^{\rm SO} = -\eta\hbsigma\cdot\smlb{\nabla V_R \times i\nabla}$ with $\hbsigma = (\hsigma_x,\hsigma_y,\hsigma_z)$ the Pauli matrices vector and $\eta$ the spin orbit coupling parameter for the surface scattering. The spin orbit coupling due to the impurity potential $V_I$ is neglected because we concentrate on the films using metals that have very weak bulk spin orbit coupling, such as Cu or Al. The full Hamiltonian now becomes 
\begin{equation}
H = H_0 + U \qwith U = V_R + V_I + V_R^{\rm SO},
\label{eqn:H}
\end{equation}
where $H_0 = \bp^2/2m^* + V_d(z)$ describes a film of constant thickness $d$ and $U$ is treated perturbatively.  

For simplicity, we assume the confining potential $V_d(z)$ takes the particle-in-box potential, and the eigenstates of $H_0$ for a thin film of thickness $d$ is $E_{n\bq} = \hbar^2(q^2+k_n^2)/2m^*$ and $\ket{n\bq} = \sqrt{2/Ad} \sin(k_n z)e^{i\bq\cdot\brho}$,
where $k_n = n\pi/d$ with $n$ the index for the transverse mode, $\bq$ is the in-plane wave-vector, and $A$ is the area in the lateral direction. Due to the scattering potential from the bulk impurities $V_I$ and surface roughness $V_R$, the $\ket{n\bq}$ state is mixed with other $\ket{n'\bq'}$ states. By Born approximation, the scattered state
\begin{equation}
\ket{n\bq^+} 
= \ket{n\bq} + \sum_{n'\bq'}\ket{n'\bq'}{\bra{n'\bq'}V_R+V_I\ket{n\bq}\ov E_{n\bq}-E_{n'\bq'}+i\epsilon}. 
\label{scstate} 
\end{equation}


The electron relaxation time, or the inverse of scattering rate, can be calculated from the transition probability $P(n'\bq',n\bq) = \abs{\avg{\bra{n'\bq'}T\ket{n\bq}}}^2$ with $T = U + U(E-H)^{-1}U$, where the double $\avg{\avg{\cdots}}$ denotes the expectation value average over the scattered state $\ket{n\bq^+}$ and the ensemble average over roughness profiles (and/or impurity distributions). Assuming that the surface roughness and the impurity distribution are uncorrelated, to the leading order in $\Lambda$ and $V_{\rm imp}$, the scattering rate is the sum of the rate from surface scattering and impurity scattering: \cite{tesanovic_quantum_1986,trivedi_quantum_1988}
$\tau^{-1}_n = \tau'^{-1}_n + \tau_0^{-1}$. Here $1/\tau'_n = n^2/\tau'$ is the channel dependent surface scattering rate and $1/\tau_0$ is the impurity scattering rate:
\begin{equation*}
{1\ov \tau'} = {\delta^2\ov a^2}{4S\ov 3n_c^3}{E_F\ov\hbar}, \quad
{1\ov \tau_0} = \smlb{1+{1\ov 2n_c}}{n_i\ov 2\pi k_F^3}{V_{\rm imp}^2\ov E_F^2}{E_F\ov\hbar},
\label{eqn:taun}
\end{equation*}
where $n_c = \lfloor{k_Fd/\pi}\rfloor$ the total number of transverse channels and $S = 3\sum^{n_c}_{n'=1} {n'^2/n_c^3} \simeq 1$. 

The in-plane velocity operator is calcualted from the Hamiltonian \Eq{eqn:H} as \cite{wang_spin-hall_2013}
\begin{equation}
\hbv_\| = -{i\ov\hbar}[\brho,H]
= {\hbp_\|\ov m^*} + {\eta\ov \hbar}\hsigma_z\hzz\times\nabla_\| V_R, 
\label{velocity}
\end{equation}
where the second term is the anomalous velocity due to surface scattering, $\nabla_\|$ is the gradient in the in-plane $\brho$ direction. We only retain $\nabla_\|V_R$ component because the gradient in $z$ direction vanishes after ensemble and state average: $\avg{\avg{\partial_z V_R}}=0$. 

Taking into the spin part of the wave function, the scattered state becomes $\ket{n\bq s^+} \equiv \ket{n\bq^+}\ket{s}$ and $\ket{s}$ is the eigen spin state with $\hsigma_z\ket{s} = s \ket{s}$. The charge current carried by $\ket{n\bq s^+}$ is
\begin{equation}
\bj_{n\bq s} = \Avg{\brak{n\bq s^+}{e\hbv_\|}{n\bq s^+}}
= e{\hbar\bq\ov m^*} +es\alpha^{\rm sj}_n \hzz\times{\hbar \bq \ov m^*}, 
\label{eqn:jnq}
\end{equation}
where $\alpha_n^{\rm sj} = m\eta/\hbar\tau'_n$ is the channel dependent dimensionless coupling parameter of side jump. The first term in \Eq{eqn:jnq} is the normal charge current, and the second term is due to the anomalous velocity and gives rise to the spin Hall current. $\alpha_n^{\rm sj}$ depends only on $\tau'_n$ but not $\tau_0$ because we only consider the spin orbit interaction $V_R^{\rm SO}$ from the surface scattering, but not from the impurity scattering.  

Besides the side jump contribution to the spin Hall current, there is usually also the skew scattering contribution due to the asymmetric distribution function, which is originated from the asymmetric transition probability $P(n'\bq',n\bq)$ between $\bq$ and $\bq'$. We found that the asymmetric transition probability vanishes, because it involves the average value of $\avg{\avg{V_RV_RV_R^{\rm SO}}} \propto \avg{\avg{\lambda_{\bq}\lambda_{\bq'}\lambda_{-\bq-\bq'}}} = 0$ according to the assumed white noise profile of surface roughness, and $\avg{\avg{V_IV_RV_R^{\rm SO}}} \propto \avg{\avg{V_I}} = 0$ when the initial and final states are different. Therefore, there is no skew scattering contribution and the coupling parameter for skew scattering vanishes: $\alpha^{\rm ss} = 0$. Physically, the vanishing of skew scattering parameter is caused by the random sign of the surface roughness potential $V_R \propto \lambda_{\brho}$, which is attractive when $\lambda_{\brho} < 0$ (or $d(\brho) > d$) and repulsive when $\lambda_{\brho} > 0$ (or $d(\brho) < d$). Therefore, the ensemble average over the surface roughness of $V_R^3$ vanishes due to the random sign. This situation is different from the impurity scattering case, in which the sign of the impurity scattering potential is fixed, \ie the impurities are all either attrative or repulsive.  


\begin{figure}[t]
\includegraphics[width=\columnwidth]{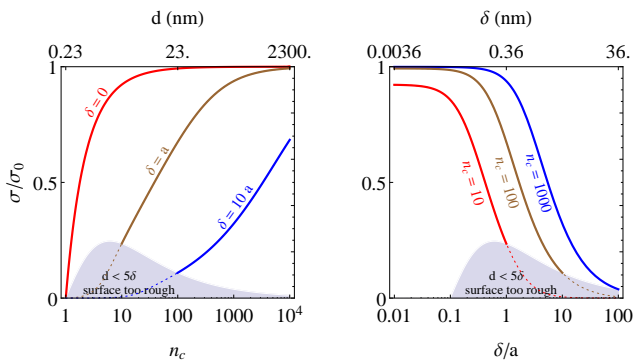}
\caption{(Color online) Longitudinal conductivity for Cu film, $k_F = 1.36\times 10^{10}$/m, $a = 3.61$\AA, and $\sigma_0 = 5.88\times10^7$~S/m ($\tau_0 \sim 24$~fs). Left: $\sigma$ as function of film thickness $n_c \simeq k_F d/\pi$ for three cases of roughness $\delta = 0, a, 10a$; Right: $\sigma$ as function of the surface roughness $\delta/a$ for three cases of film thickness $n_c = 10, 100, 1000$. In the shaded area, the surface is too rough ($\delta/d > 0.2$) and the perturbation assumption fails. }
\label{fig:sigma}
\end{figure}

When an in-plane electric field $\bE$ is applied, the Fermi circle shifts $\delta\bq_n=e\bE\tau_n/\hbar$ for channel $n$, and the corresponding non-equilibrium distribution function is 
\begin{equation}
g_{n\bq} = f_n(\bq+\delta\bq_n) - f_n(\bq) = {e\hbar\tau_n\ov m^*}\delta(E_{n\bq}-E_F)\bq\cdot\bE, 
\end{equation} 
where $f_n(\bq) = \Theta(E_{n\bq}-E)$ is the Fermi-Dirac distribution function with the Heaviside $\Theta$ function. The normal velocity term in \Eq{eqn:jnq} contributes to the in-plane longitudinal charge current 
$\bJ = \sum_{n\bq s}g_{n\bq}\bj_{n\bq s} = \sigma \bE$ with the longitudinal conductivity
\begin{equation}
\sigma = 
{3\sigma_0\ov 2n_c}\sum_{n=1}^{n_c} {\tau_n\ov\tau_0} \smlb{ 1 - {n^2\ov n_c^2}} 
\simeq \begin{cases}
{e^2k_F\ov 2\hbar}\smlb{d\ov \delta}^2, \\
\sigma_0\smlb{1-{3\ov 4n_c}},
\end{cases}
\label{eqn:sigma} 
\end{equation} 
where $\sigma_0 = {k_F^3\ov 3\pi^2}{e^2\tau_0\ov m^*}$ is the bulk conductivity. The upper (lower) approximation in \Eq{eqn:sigma} corresponds to the surface (impurity) scattering dominating case with $\tau'_n\ll\tau_0$ ($\tau'_n\gg \tau_0$). The charge conductivity \Eq{eqn:sigma} agrees with Ref. \onlinecite{trivedi_quantum_1988}. 
\Figure{fig:sigma} shows the film thickness and surface roughness dependence of $\sigma$ in \Eq{eqn:sigma}.
Because we treat the surface roughness as perturbation, \Eq{eqn:sigma} is valid only for when the roughness is small comparing to the film thickness: $\delta\ll d$. Therefore, \Eq{eqn:sigma} is not a good approximation in the shaded area in \Figure{fig:sigma}.   


\begin{figure}[t]
\includegraphics[width=\columnwidth]{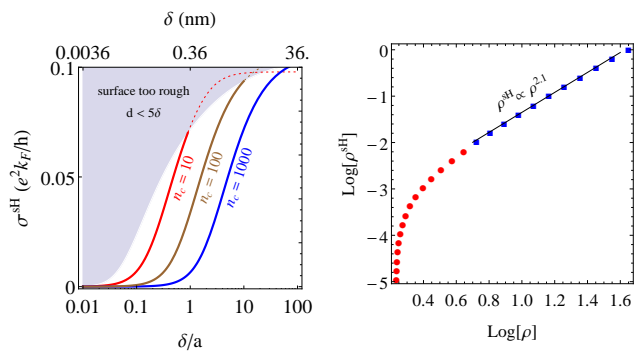}
\caption{(Color online) Spin Hall conductivity $\sigma^{\rm sH}$ for Cu film with the same parameter as in \Figure{fig:sigma} and $\bar{\eta} = 0.5$. Left: $\sigma^{\rm sH}$ as function of surface roughness $\delta/a$; Right: spin Hall resistivity $\rho^{\rm sH} = \sigma^{\rm sH}/\sigma^2$ as function of resistivity $\rho = 1/\sigma$ (by varying surface roughness $\delta$) for $n_c = 1000$. The linear fit in the surface roughness dominating region (blue squares) has $\rho^{\rm sH} \propto \rho^{2.1}$, suggesting a side jump  mechanism. }
\label{fig:sigmaSH}
\end{figure}

The second anomalous term in \Eq{eqn:jnq} does not lead to any charge current because of its opposite spin dependence, but gives rise to an in-plane pure spin current in the transverse direction or a spin Hall current 
$\bJ_s = \sum_{n\bq s}sg_{n\bq}\bj_{n\bq s} = \sigma^{\rm sH} \hzz\times\bE$. In the unit of charge current, the spin Hall conductivity
\begin{align}
\label{eqn:sigmasH} 
&\sigma^{\rm sH} = {e^2 k_F\ov h}{\bar{\eta}\ov \pi}
\sum_{n=1}^{n_c} {\tau_n\ov\tau'_n} \smlb{ {1\ov n_c} - {n^2\ov n_c^3}} 
\simeq {e^2k_F\ov h} {\bar{\eta}\ov \pi}
\begin{cases} 
{2\ov 3}, \\
{2n_c^2\tau_0\ov 15\tau'},
\end{cases}
\end{align} 
with $\bar{\eta} = \eta k_F^2$ being the dimensionless spin orbit coupling parameter. 
The upper (lower) approximation in \Eq{eqn:sigmasH} corresponds to the surface (impurity) scattering dominating case with $\tau'_n\ll\tau_0$ ($\tau'_n\gg \tau_0$). The spin Hall conductivity in \Eq{eqn:sigmasH} for surface roughness dominating case is independent of the surface roughness $\delta$ due to the following reason: $\sigma^{\rm sH}$ is proportional to both the side jump coupling parameter $\alpha_n^{\rm sj}\propto \tau'^{-1}$ and the Fermi circle shift $\delta\bq_n\propto \tau_n\sim \tau'$, thus no dependence on the relaxation time $\tau'$ or the surface roughness. This behavior is the same as the side jump contribution to the spin Hall conductivity in bulk materials. \cite{takahashi_spin_2008} The left panel of \Figure{fig:sigmaSH} plots $\sigma^{\rm sH}$  as the function of $\delta$, showing that $\sigma^{\rm sH}$ is larger for more rough surface and thinner film. In conventional bulk spin Hall effect, the relation between the spin Hall resistivity $\rho^{\rm sH} = \sigma^{\rm sH}/\sigma^2$ can be expressed in terms of the longitudinal resistivity $\rho = 1/\sigma$: $\rho^{\rm sH} = a \rho + b \rho^2$, where the linear and quadratic terms are due to the skew scattering and side jump mechanisms, respectively. We show in the right panel of \Figure{fig:sigmaSH} the log-log plot of $\rho^{\rm sH}$ as function of $\rho$, which includes the bulk resistivity $\rho_0 = 1/\sigma_0$ and the surface scattering induced resistivity $\rho-\rho_0$. In the surface scattering dominating region (blue squares), the slope of a linear fit is 2.1, \ie $\rho^{\rm sH} \propto (\rho-\rho_0)^{2.1} \simeq \rho^{2.1}$, suggesting that the surface roughness induced spin Hall effect is due to the side jump mechanism. The slight deviation from slope 2 is due to the mixing of bulk relaxations, as well as the different relaxation time $\tau_n$ for different transverse channels.  

\begin{figure}[t]
\includegraphics[width=\columnwidth]{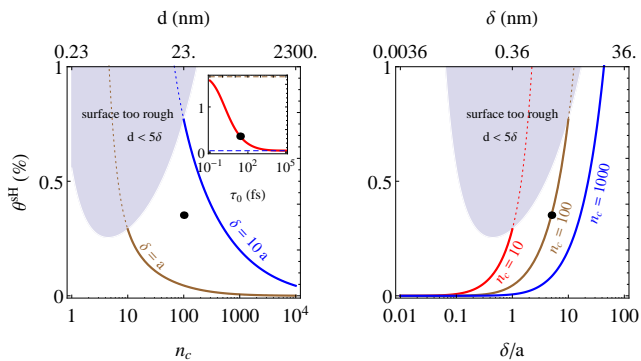}
\caption{(Color online) The spin Hall angle (in percent) for Cu film with the same parameter as in \Figure{fig:sigma} and $\bar{\eta} = 0.5$. Left: as function of film thickness $n_c \simeq k_F d/\pi$; Inset figure shows the $\theta^{\rm sH}$ dependence on the bulk relaxation time $\tau_0$ for $n_c = 100$ and $\delta = 5a$ with the dashed lines given by the limitting values in \Eq{eqn:thetasH}; Right: as function of the surface roughness $\delta/a$. The black dot in all plots corresponds to the same point with $n_c = 100$ ($d \simeq 23$~nm), $\delta = 5a \simeq 1.8$~nm, $\tau_0 \sim 24$~fs for Cu with conductivity $\sigma_0 = 5.88\times 10^7$~S/m, and has spin Hall angel $\theta^{\rm sH} \simeq 0.35\%$. }
\label{fig:thetaSH}
\end{figure}

With both the longitudinal conductivity \Eq{eqn:sigma} and the spin Hall conductivity \Eq{eqn:sigmasH}, the spin Hall angle is calculated as,
\begin{equation}
\theta^{\rm sH} = {\sigma^{\rm sH}\ov \sigma} 
\simeq \bar{\eta} \smlb{\delta\ov d}^2 \begin{cases}
{1\ov 3\pi^2}, \\
{n_c\ov 30}\smlb{2\pi\ov k_Fa}^2.
\end{cases}
\label{eqn:thetasH} 
\end{equation} 
The upper (lower) approximation in \Eq{eqn:thetasH} corresponds to the surface (impurity) scattering dominating case with $\tau'_n\ll\tau_0$ ($\tau'_n\gg \tau_0$).
As shown in \Figure{fig:thetaSH}, the spin Hall angle can be enhanced by i) decreasing film thickness $n_c$, ii) increasing surface roughness $\delta$, iii) decreasing bulk relaxation time ($\tau_0$) or increasing bulk resistivity ($1/\sigma_0$). For $n_c = 100$ and $\delta = 5a$, the spin Hall angle for thin films made of Cu, Al, and Ag are listed in \Table{tab:shangle}. For Cu, $d \simeq 23$nm and $\delta \simeq 1.8$nm, the spin Hall angle for Cu film can reach a fraction of a percent ($0.35\%$), which is comparable to that for bulk Au. 

For comparison, we also carry out the same calculation for a thin film where the spin orbit interaction comes from the impurity scattering instead of the surface scattering, \ie the $V_{\rm R}^{\rm SO}$ in $H$ \Eq{eqn:H} is replaced by $V_I^{\rm SO} = -\eta_I \hbsigma\cdot(\nabla V_I \times i\nabla)$ with $\eta_I$ the spin-orbit interaction parameter for impurity scattering potential $V_I$. 
Following Takahashi \textit{et al}, \cite{takahashi_spin_2008} we find the coupling parameter of side jump and skew scattering as
\begin{equation}
\alpha^{\rm sj}_I = {m^*\eta_I\ov \hbar\tau_0}, \qand
\alpha^{\rm ss}_I
= {\bar{\eta}_I\ov 12\pi}{V_{\rm imp}\ov E_F}{\sigma\ov \sigma_0}
\label{eqn:alphai} 
\end{equation}
where $\bar{\eta}_I = \eta k_F^2$ and $\sigma$ is the same conductivity as \Eq{eqn:sigma}.
The spin Hall conductivity $\sigma_I^{\rm sH} = (\alpha^{\rm sj}_I + \alpha^{\rm ss}_I )\sigma$.
Therefore the spin Hall angle for the bulk impurity induced spin Hall effect in metallic thin film is $\theta_I^{\rm SH} = \alpha^{\rm sj}_I + \alpha^{\rm ss}_I$. Since side jump contribution $\alpha^{\rm sj}_I$ is a constant, independent of film thickness and/or surface roughness, and skew scattering contribution $\alpha^{\rm ss}_I$ has the same parameter dependence as the longitudinal conductivity $\sigma$, therefore the spin Hall angle by impurity scattering in metallic thin films decreases with decreasing film thickness and/or increasing surface roughness (following the same trend as $\sigma$ in \Figure{fig:sigma}). This behavior is opposite to that for the spin Hall angle by surface scattering (see \Figure{fig:thetaSH}). Therefore, it is possible to distinguish the origin of the spin Hall effect from the thickness and/or surface roughness dependence of the spin Hall angle.  

\begin{table}[t]
\centering
\begin{tabular}{c|c|c|c}
	\hline
	Material& $\sigma_0$ ($10^7$S/m)& $k_F$ (1/\AA) & $\theta^{\rm sH}$ \hhline
	Cu 	& $5.88$		& $1.36$	& $0.35\%$ \\
	Ag 	& $6.21$		& $1.19$	& $0.32\%$ \\
	Au 	& $4.55$		& $1.21$	& $0.37\%$ \\
	Al 	& $3.65$		& $1.75$	& $0.48\%$ \\ \hline
\end{tabular}
\caption{Surface roughness induced spin Hall angle for thin films of several normal metals that has no bulk spin Hall effect. In all cases, $n_c = 100$, $\delta = 5a$, $\bar{\eta} = 0.5$, and electron effective mass equals to the free electron mass $m^* = m$. Data for $\sigma_0, k_F$ from \cite{ashcroft_solid_1985,kittel_introduction_2005} } 
\label{tab:shangle}
\end{table}

To realize the surface roughness induced spin Hall effect experimentally, it is necessary to have two crucial ingredients simultaneously: i) the surface roughness, which acts as surface impurities, ii) the strong interfacial (not bulk) spin-orbit interaction. The former can be controlled by thin film growing process or the surface polishing technique. The latter is possible by coating the metallic thin film surface with materials with large spin orbit interaction, such as Pt or oxides with heavy elements. In such a way, one may utilize the strong scattering due to roughness and the strong spin-orbit interaction from the heavy elements. 

In conclusion, we predict that, in metallic thin films without bulk spin-orbit interaction, the spin Hall effect can be realized by surface roughness. For Cu film with sizable interfacial spin-orbit interaction, the spin Hall angle can be as large as a fraction of a percent ($0.35\%$), comparable to that in Au. 

This work was supported by the special funds for the Major State Basic Research Project of China (2014CB921600, 2011CB925601) and the National Natural Science Foundation of China (91121002). 

\bibliographystyle{apsrev}  
\bibliography{prl}

\end{document}